# Feature Selection for Generator Excitation Neurocontroller Development Using Filter Technique

Abdul Ghani Abro [1], Junita Mohamad Saleh [2]

**School of Electrical & Electronics Engineering**
**Engineering Campus, Universiti Sains Malaysia**
**14300 Nibong Tebal, Seberang Perai Selatan**
**Penang, Malaysia**

**Abstract**

*Essentially, motive behind using control system is to generate suitable control signal for yielding desired response of a physical process. Control of synchronous generator has always remained very critical in power system operation and control. For certain well known reasons power generators are normally operated well below their steady state stability limit. This raises demand for efficient and fast controllers. Artificial intelligence has been reported to give revolutionary outcomes in the field of control engineering. Artificial Neural Network (ANN), a branch of artificial intelligence has been used for nonlinear and adaptive control, utilizing its inherent observability. The overall performance of neurocontroller is dependent upon input features too. Selecting optimum features to train a neurocontroller optimally is very critical. Both quality and size of data are of equal importance for better performance. In this work filter technique is employed to select independent factors for ANN training.*

**Keywords:** *neural network, mlp, feature selection, regression analysis, generator excitation*

## 1. Introduction

In recent years it has been recognized to impart more flexible control systems, it is necessary to incorporate other elements, such as course of thoughts, reasoning and heuristics into algorithmic techniques of conventional adaptive and optimal control theory. For proper designing of adaptive controller flexibility is main characteristic to incorporate and Artificial Neural Network (ANN) offers highly flexible structure. The use of an ANN with its learning ability avoids complex mathematical analysis in solving control problems when plant dynamics are unpredictably complex and highly non-linear [1]. This is a distinctive advantage over the traditional non-linear control methods.

ANNs are parallel distributed processing systems capable of synthesizing a complex and highly nonlinear mapping from input feature space to output space [2]. The parallel processing element distribution not only gives higher degree of tolerance but also the capability of fast information processing. Another important feature of ANN is learning and adaptation. A well trained ANN has the ability to generalize training pattern. In addition to their ability to produce high quality results for large, noisy or incomplete data sets, ANNs have been found effective in identifying patterns and other underlying data structures in multidimensional data [3].

Importance of input variables is evident as the input vector needs to capture all characteristics of complex functions. Features, variables, attributes, parameters are used alternatively for input vectors given to ANN. Feature selection is a problem of selecting the subset of features that is needed to describe the target concept in a give data set, indeed [4-6]. Keeping this in mind, one can say that it is not necessary that best individual feature correspond to best set of feature. Therefore, for best subset of features researcher better undergo all possible combinations of features available in feature set for optimum efficiency.

Feature selection is fundamental because it allows us to reduce the various effects causing information overlapping, noise induction, highly complex computation, cost of computation, memory requirement, time to compute and inter variable correlation [7]. Alternatively, too few features may carry very low content information and too many may cause irrelevant features, complex mapping and data over fitting [3]. However, it should be pointed out that a larger number of training data should always be favored as opposed to smaller number. Hence, it is issue of harmonizing irrelevant data and information.

There are two techniques employed for feature selection. One is filter-based approaches employing statistical tests for feature selection. Another is called wrappers methods exploit the knowledge of the specific structure of the learning algorithm and cannot be separated from it [8]. In the absence of valid and reliable evaluation, there currently exists no consensus on which methodology should be applied under which data condition [9]. Wrapper technique based feature selection is computationally very expensive [7] and this is big flaw behind not being used so frequently. In comparison filters based feature selection methodology is most frequently used for efficient feature evaluation [9]. In this research work forward selection based on statistical methods is used to select optimum sub set of features.

Since the discovery of Multi Layer Perceptrons' (MLP) nonlinear problem solving ability, there has been an explosive growth in application of ANN into control problems. MLP is most commonly used ANN topology and type is a type of feed-forward network. MLPs are finding more applications because of their simplicity and requirement of lesser features to approximate any function up to same degree of certainty. It contains one or more hidden layers. The number of nodes in the hidden layers defines the complexity and the power of the neural network model to describe underlying relationship and structure inherent in training data [3]. ANNs have a specific nonlinear function associated with every number of hidden layer size. However, it cannot be interpreted because of poor interpretability of ANN. Generally, one hidden layer with sigmoidal activation function is used with sufficient accuracy to approximate any nonlinear function. A key challenging aspect of the MLP-ANN is the optimization of network training protocols that include network architecture and training stopping criterion [10].

Due to the nonlinear and highly dynamic nature of power system and complexity involved in realization of optimal and nonlinear controllers, artificial intelligence particularly ANNs are finding wide variety of applications in operation and control of power system [11-17]. There are thousands of papers published in this area but literature reviewed here is as per scope of this paper focusing on excitation of synchronous generator. Work proposed by [18] has used Functional Link Net (FLN) and technique researched in [19] has proposed a method equivalent to conventional self tuning adaptive control utilizing RBF feedforward network. Research proposed in [1], [20] and [21] is based on indirect adaptive control, utilizing three layer MLP to realize model and neurocontroller. Adaptive Critic Design (ACD) based control utilizes Hamilton–Jacobi–Bellman equation based optimal control algorithm. Duel Heuristic Programming (DHP) based ACD has been shown to perform better [22]. Work proposed by [23] use MLP based critic control, whereas MLP and RBF based critic control comparison was carried out in [24]. RBF showed better performance for low magnitude disturbances. In [25] by using RBF based adaptive critic neurocontroller, it is showed that performance of neurocontroller is better even when conventional excitation system is equipped with power system stabilizers (PSS).

In indirect adaptive based control, since link between current system state and the controller parameters are totally ignored [26] and the identified model has error of considerable percent [27], then may controller generate erroneous signal and lead to oscillatory response. Whereas in ACD, complicated control algorithm needs more computational time to calculate control signal [28] and response time is key to close loop control system performance specifically dynamic system such as power system. Additionally, reliability of ACD based control loop is also low. On the contrary, if not impossible at least it is time-consuming to train neurocontroller offline for every operating condition.

Generalization means to capture trend in data instead of fitting every training data set. Alternatively, close inputs ought to generate close outputs. For better generalization early stopping criterion plays very important role. Apart from that, early stopping of ANN training saves training time. However, if criterion to stop training is not appropriately chosen that may lead to under trained network. In this research work, ANN training was stopped on the basis of the network's performance on validation data set. This early stopping criterion is not used in the realm of power system control and operation. This is explained in the last section.

As explained, a highly challenging task to train ANN for power system control and operation is selection of input features. Aforementioned literature review reveals that variables given in Table 1 were used for ANN training to control excitation of synchronous generator. The output of the excitation system is called excitation voltage ($V_f$) and it is a dependent parameter. Detailed explanation of excitation system's impact on generator operation is given in next section.

Table 1 Input feature used in ANN training

| | |
|---|---|
| $\Delta V_T$ | Terminal Voltage |
| $\omega$ | Rotor Speed |
| P | Active Power |
| Q | Reactive Power |

$\Delta V_T$ is deviation of terminal voltage from reference voltage i.e. $V_{REF} - V_T$

No proper procedure has been reported for selecting input features for generator excitation neurocontroller training. More input features may require many processing elements and hence more information processing time. On the other hand, multicollinearity between input features may inhibit a neurocontroller's learning capability. In addition, un-correlated input and output space make the mapping very complex. Furthermore, generator terminal voltage can be sensed by different combinations of Table 1 parameters and even with few additional factors. This analysis is first of its own kind, to the authors best knowledge such analysis was not carried keeping statistical and engineering constraint both at a time.

Statistical methods and ANN have been used for prediction and approximation, with ANN giving higher accuracy in high dimensional problems. In fact, the most commonly used ANN topology, called multilayer perceptron is nothing more than nonlinear regression. By using statistical methods optimum parameters can be found [29] for enhancing neuro controller learning capability while generator dynamics remain unaffected. Objective of this paper is to generate optimal set of training features and to compare performance of statistical regression and ANN.

This paper is divided into three parts. The immediate section discusses the model considered for data generation. Second segment describes data analysis based on statistical methods and last part focuses on ANN output and comparison.

## 2. Power System Modeling

Power system is spread over very wide region from one end of country to another end and sometime from one continent to another, comprising of generation, transmission and distribution sections. The primary element of generation section is synchronous generator also termed as alternator. Synchronous generator consists of stator called armature and rotor also known as field. Field is responsible for keeping air gap magnetic flux constant leads to constant terminal voltage. The key to proper operation of synchronous generator is maintenance of synchronism between rotating armature flux and revolving field flux. The strength of synchronism largely depends upon the strength of air gap magnetic flux and alternatively dependent upon excitation system performance. Synchronism can be jolted by faults induced anywhere in a power system, but the extreme disturbance is fault introduced at terminals of a generator. Fault deteriorates the strength of magnetic flux as explained by armature reaction phenomenon and so has the effect on synchronism. The mechanical angle between rotor magnetic field and armature magnetic flux of a generator is known as the load angle or power angle, δ.

The ability of power system to regain a state of operating equilibrium after being subjected to a physical disturbance or fault is called power system stability. In addition, neither a unit at generating station nor a portion of power system should lose synchronism with respect to the generating station or the power system [30]. Power system stability enhancement has captured growing attention of researchers in recent times after occurrence of major blackouts [31].The excitation system's output is based on the difference (ΔV) between reference voltage and terminal voltage. The fault causes a decrease in air gap flux density, depending upon the direct and quadrature axis sub-transient and transient time constant. Moreover duration of fault, and decrease in terminal voltage have great influence on air gap flux density reduction. This leads to increase in ΔV, so the output of generator excitation will shoot up to compensate error. Stability may be enhanced by rapidly increasing excitation current [32]. ANN requires quite considerable time to tune weights but it is fast and accurate once tuned properly. In this research work besides variables given in Table 1, the effect of one more new variable, deviation of quadratic voltage from reference voltage i.e. $\Delta V_q = V_{ref} - V_q$ was analyzed on excitation voltage ($V_f$). Terminal voltage is the vector sum of direct and quadratic voltage components. Quadratic voltage was preferred over direct voltage because of higher correlation constant. The reference value for quadratic voltage is achieved by putting one instead of terminal voltage in equation combining direct, quadratic and terminal voltage.

Power system stability enhancement is referred to reducing risk of losing stability by inserting additional signals into the system to smooth out the system dynamics. During steady state excitation system should be driven by only voltage difference. Contrastingly, during transient state rotor swings ΔV undergoes oscillations caused by change in rotor angle. It is compulsory to add additional information to neurocontroller for damping out oscillations. Rotor speed, active power or both are usually used variables for generating stabilizing signals [11, 33, 34]. In this research work one more parameter load angle (δ) is also included for analyzing learning performance based on its correlation with excitation voltage and active power. Selection of load angle will not affect negatively the generator dynamics because active power and load angle are proportional as evident from equation

$$P = \frac{E_f * V_T}{X_S} * \sin \delta \qquad (1)$$

where P is active power, $E_f$ is internal generated voltage, $V_T$ is terminal voltage, $X_S$ is synchronous reactance of generator and δ is load angle. Single machine infinite bus system (SMIB) power system model is considered for generating data, as shown in Figure 1. This model simulates a generator connected with the rest of power system.

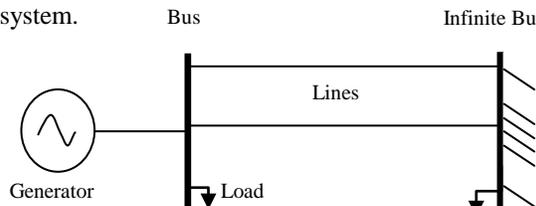

Figure 1 A single machine-infinite bus system

Simulation of the model was carried out on Matlab/Simulink Toolbox with generator rating 13.8KV,

150MVA, 50Hz at load (0.09+j0.056) Ω. The generator parameters and excitation system parameters are given in Table 2 and 3 respectively. A three phase to ground fault was simulated to analyze system transient stability. Figure 2 and 3 show terminal voltage and load angle behavior after simulation of 120ms fault at generator terminals. Both figures depict stable behavior of the generator. This implies data were collected from a stable system.

Table 2 SYNCHRONOUS GENERATOR PARAMETERS

| $X_d$ = 1.83 | $X_q$ = 1.7 | $R_{Stator}$ = 0.003 |
|---|---|---|
| $X'_d$ = .24 | $X'_q$ = 0.43 | Inertia = 3.6 |
| $X''_d$ = .20 | $X''_q$ = 0.26 | Hz = 50 |
| $T'_d$ = 0.3s | $T''_d$ = 0.04s | $T''_q$ = 0.031s |

Table 3 EXCITATION PARAMETERS

| Ka = | 2.50 | Ta = | 0.001s |
|---|---|---|---|
| Ke = | 1.5 | Te = | 0.3s |
| Kf = | 1 | Tf = | 0.003s |

The field of statistics deals with the collection, presentation, analysis and use of data to make decisions. Statistical methods are used to assist for describing and appreciating variability. Variability means the successive observations of a system or phenomenon do not always produce exactly the same result. Hence statistical thinking gives us a useful way to incorporate variability into decision making process.

Statistical analysis was carried out using Minitab software. In statistical modeling data generation play an important role in model acceptance. In this work aforesaid model simulation include ±10% change in Vref, self clearing and not self clearing three phase to ground fault at generator terminals and transmission line tripping and addition as the types of disturbances. Data were sampled at 200Hz sampling frequency. Then fifty random samples were taken for further analysis. Care was taken that the sample should be a true representation of whole population space

## 2. Data Analysis

Since power system requires high degree of reliability, hence fewer inputs are preferred to use. The efficiency of control signal is increased by using time delayed values as power system is dynamic system. This is the main reason why this analysis did not use stepwise regression analysis and only relied on linear regression. The statistical modeling process involved three steps [35]: (i) correlation analysis, (ii) regression analysis, and (iii) model assessment [36]. These steps are discussed below.

*(a) Correlation* is the process for determining the strength of relation between dependent and independent variables. Table 4 shows Pearson correlation between various independent parameters and the dependent factor, excitation voltage ($V_f$). The table also shows significance, called probability value (P-value). Pearson correlation was chosen because the data is scaled type, i.e. value varies from -∞ to +∞.

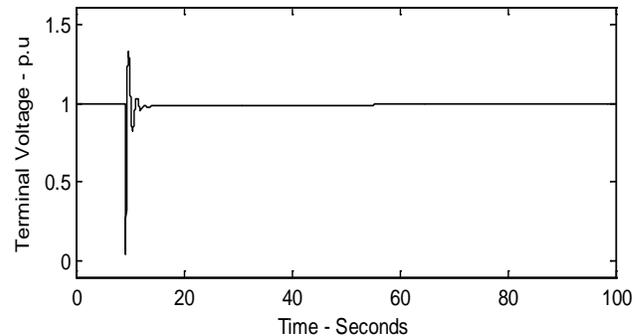

Figure 2 Terminal voltage after 120ms fault at generator terminals

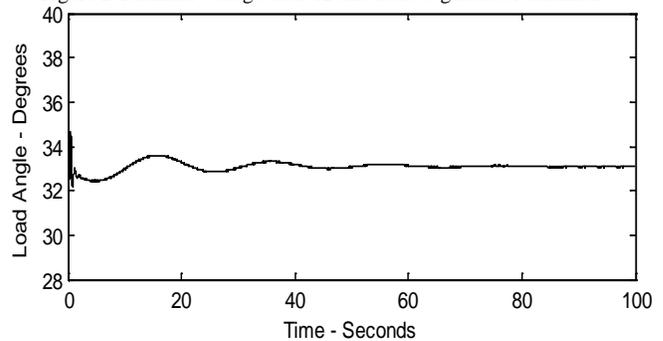

Figure 3 Load angle (δ) transition after fault

Table 4. Statistical Correlation Test Output

| Independent Variables | Correlation coefficient | Significance P-Value |
|---|---|---|
| $\Delta V_T$ | 0.587 | 0.000 |
| ω | -0.06 | 0.486 |
| P | 0.648 | 0.000 |
| Q | 0.635 | 0.000 |
| $\Delta V_q$ | 0.759 | 0.000 |

Correlation is between Excitation Voltage ($V_f$) and parameters given.

For 95% confidence level, the significance of less than 0.05 considered statistically meaningful. The table suggests strongest correlation between quadratic voltage ($V_q$) and excitation voltage ($V_f$), followed by active (**P**) and reactive power (**Q**). The results suggest no relationship between rotor speed and excitation voltage and this parameter is currently being used as auxiliary stabilizing signal. Keeping in view correlation results rotor speed is eliminated from further analysis.

(b) *Regression Analysis* gives the prediction of dependent variable based on independent factors of an empirical model. The regression equation is given as below [37],

$$\Upsilon = \beta o + \sum_{i=1}^{n} \beta i X i + \varepsilon \quad (2)$$

where β are constants and X is the independent variable and Y is the dependent variable. The random error term ε is assumed to have zero mean, constant variance $\sigma^2$ and normally distributed [38].

The accuracy of regression model is determined by coefficient of multiple determination i.e. $R^2$. Higher $R^2$ value indicates good prediction accuracy of model. Nevertheless, using only $R^2$ is not always a good indicator of model adequacy. Seeing that $R^2$ increases with addition of another regressor variable irrespective of whether additional variable is statistical significant or not. The adjusted coefficient of multiple determination i.e. $R^2$(Adj) is a better reflection of the model adequacy along with $R^2$. $R^2$(Adj) will increase only when additional factor is statistically significant. Lower standard deviation (S) is also conceived an indicator for better performance of the model. The low S means data set tends to be very close to mean and assumed mean in regression definition is zero.

Table 5 gives the regression analysis results of the models without having stabilizing signal. Model 1contains ΔVq and model 2 comprises of $\Delta V_T$. Coefficient of multiple determination ($R^2$) of model 1 is higher than model 2. Therefore, it can be concluded that $\Delta V_q$ has higher prediction accuracy than $\Delta V_T$. Hence it is expected, neuro controller trained on model 1 may give less error than model 2. Value of $R^2$(Adj) is higher for model 1 than model 2. Since in this analysis each model contains only one element $R^2$(Adj) does not serve its purpose here. However, this value is given here to compare in next step when stabilization signals are added and compared in different combinations.

Nonetheless, only high gain excitation system can produce low frequency oscillations which ultimately lead to unstable system. Therefore, auxiliary stabilizing signals addition to excitation system is essential for stability enhancement [39]. The following explanation considers the models comprising of additional signals. Table 6 describe the regression output of models containing active power (P) and reactive power (Q) as stabilizing input feature in combination to voltage deviation signals.

After contemplating Table 6 it can be concluded that not only the additional signals stabilize the system but also increase prediction accuracy. With additional signal addition $R^2$(Adj) is also increased, which is another evidence to believe higher prediction accuracy of Table 6 models. However, authors are reluctant to select P and Q as stabilizing parameter owing to higher VIF factor. Variance Inflation Factor (VIF) predicts correlation among predictors. From statistical analysis perspective VIF value up to 10 is considered normal. Nevertheless, lesser VIF value means better performance, since ANNs output is highly sensitive to VIF value. In addition, active power and reactive power being electrical signals have lower response time. These quantities are very sensitive to noise in comparison to mechanical signals conceived here onward.

Table 5 Regression output of Models without Stabilizing Signals

|  | Model 1 | Model 2 |
|---|---|---|
| Constant | 1.879 | -0.214 |
| $\Delta V_q$(coefficient) | 24.944 | - |
| Significance | 0.000 | - |
| $\Delta V_T$(coefficient) | - | -35.617 |
| Significance | - | 0.000 |
| S | 0.95 | 1.233 |
| $R^2$ | 73.3 | 59.8 |
| $R^2$(Adj) | 70.2 | 56.1 |

Table 6 Performance index for Regression of Models with Stabilizing Signals

|  | Model 3 | Model 4 | Model 5 | Model 6 |
|---|---|---|---|---|
| Constant | 2.718 | 7.140 | 4.490 | 3.953 |
| $\Delta V_q$(Coefficient) | 22.441 | - | 17.982 | - |
| Significance | 0.000 | - | 0.000 | - |
| VIF | 5.6 | - | 6.9 | - |
| $\Delta V_T$(Coefficient) | - | 9.653 | - | 19.620 |
| Significance | - | 0.251 | - | 0.021 |
| VIF | - | 9.8 | - | 8.3 |
| P (Coefficient) | - | 8.151 | 3.525 | - |
| Significance | - | 0.001 | 0.071 | - |
| VIF | - | 6.4 | 7.8 | - |
| Q (Coefficient) | 22.441 | - | - | 17.879 |
| Significance | 0.493 | - | - | 0.03 |
| VIF | 5.7 | - | - | 4.9 |
| S | 0.985 | 1.108 | 0.955 | 1.1839 |
| $R^2$ | 75.5 | 69.0 | 77.2 | 64.7 |
| $R^2$(Adj) | 73.9 | 67.0 | 75.5 | 62.3 |

Table 7 also shows regression analysis output but with different stabilizing signal in combination to voltage. Higher prediction accuracy and low VIF value prophecy better performance of Table 7 models than Table 6 models. Models 7 and 8 consisting of load angle (δ) in combinations of terminal voltage (Vt) and quadratic voltage (Vq). $R^2$ of model containing quadratic voltage is higher than model comprising terminal voltage and value of $R^2(Adj)$ is also higher. The S of model 8 is lower than model 7 too, shown in Table 7. However, VIF value of model 7 and model 8 is almost equal. It can be deduced from Table 7 figures, set of input parameters containing quadratic voltage has higher prediction accuracy than set consisting of terminal voltage. Hence it is anticipated, neurocontroller trained on model 8 may give less error than model 7.

TABLE 7 Performance index for Regression of Models with Stabilizing Signals

|  | Model 7 | Model 8 |
|---|---|---|
| Constant | -1.825 | -9.230 |
| $\Delta V_q$(Coefficient) | - | 29.697 |
| Significance | - | 0.000 |
| VIF | - | 2.4 |
| $\Delta V_T$(Coefficient) | 36.522 | - |
| Significance | 0.000 | - |
| VIF | 2.9 | - |
| $\delta$(Coefficient) | 0.0419 | 0.2995 |
| Significance | 0.744 | 0.005 |
| VIF | 1.6 | 1.5 |
| S | 1.2451 | 0.9073 |
| $R^2$ | 60.9% | 79.2% |
| $R^2$(Adj) | 58.3% | 77.9% |

(c) *Model Assessment* step is carried out after regression model is developed. Acceptability and reliability of model are carried out in this step. Fitting a regression model requires few assumptions, meeting them tells credibility of the model. It is assumed while fitting data that the residuals are randomly distributed and lie within ±2. The residuals of a regression model are given by

$$e = y_{des} - y_{est} \quad (3)$$

where e is the error, $y_{des}$ is the desired output and $y_{est}$ is the estimated output. Analysis of residuals is helpful in checking assumption that the errors are approximately normally distributed with constant variance. For 95% confidence interval more than or equal to 95% residuals of model ought to lie within ±2 range [36].

Performance analysis of models 3 to 6, using steps suggested in model assessment depicts pretty poor picture too. Therefore, their assessment is not shown here. Only assessment comparison of Table 7 models is described here.

Model adequacy is analyzed by beholding the behavior of model residuals. Figure 4, 5, and 6 are related with residuals of model 8 whereas Figure 7, 8 and 9 are associated with residuals of model 7. Residual plots comparison, Figure 4 and 7, of both models exhibit that the residual distribution of both model is satisfactory. The points at (0.72,-4.5) and (1.29,-0.2) are not outliers, but these represent one of many different disturbances simulated during data generation phase. More than 95% residuals of model 8 lie in the range of ±2, which indicates that the assumptions of randomly distributed residual is satisfied, as shown by Figure 5 and 6. Whereas less than 95% of residuals of model 7 lie within range of ± 2, as indicated by Figure 8 and 9.

The distribution of residuals along regression fit is shown in Figure 5 and 8 for model 8 and model 7, respectively. The comparison of both plots expose that distribution of model 8 residuals is approaching normality more than model 7 residuals.

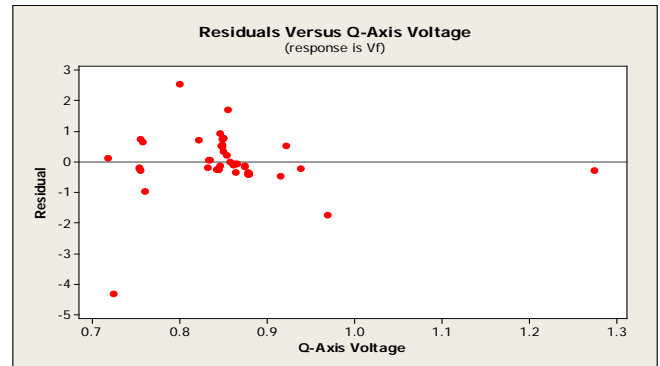
Figure 4 Residual plots of Model 8

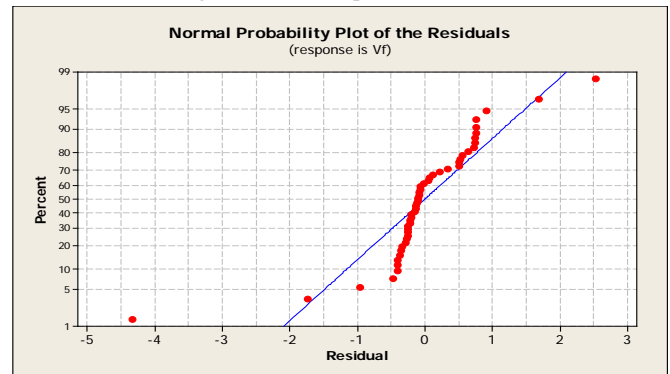
Figure 5 Normal distribution of residual of Model 8

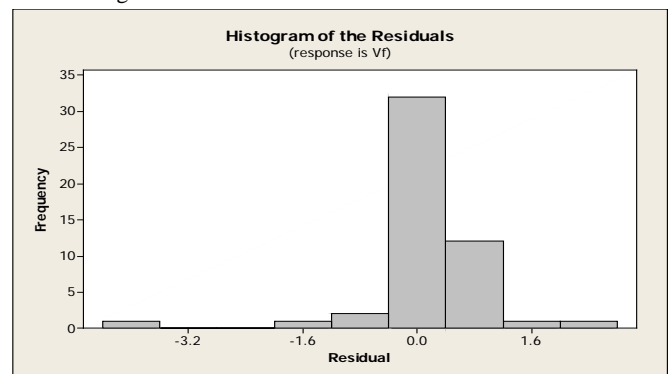
Figure 6 Histogram of Model 8 residuals

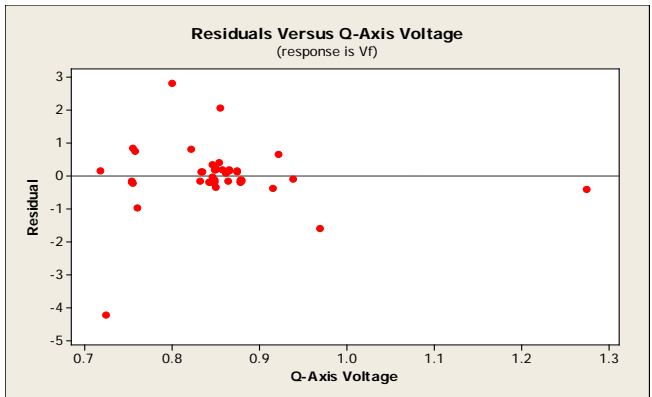

Figure 7 Residual plots of Model 7

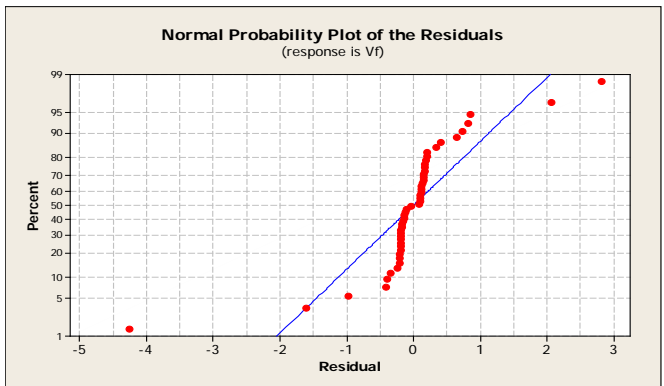

Figure 8 Normal distribution of residual of Model 7

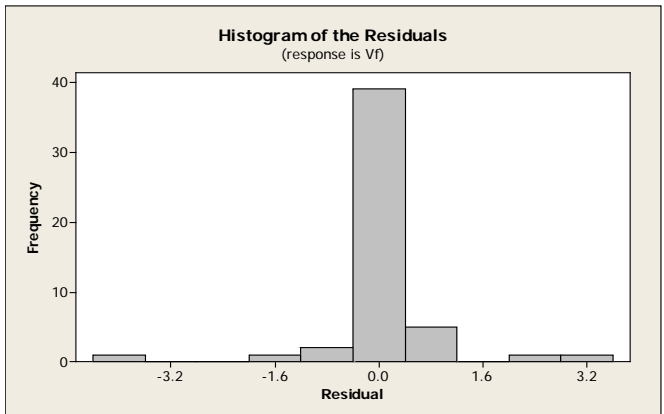

Figure 9 Histogram of Model 7 residual

## 4. Artificial Neural Network

In this section ANN output is discussed. In this research work MLP was chosen because of its simplicity and it is most usually used neural network [35]. A perceptron network with its adjustable hidden layer values is nonlinearly parameterized. MLP was trained using Levenberg-Marquardt Error Back Propagation. A highly challenging characteristic for a trained ANN is how well it performs when presented with new data i.e. generalization

Advantages of generalization lie in adaptability, fault tolerance and model-free estimation by constructing input output mapping. As discussed in introduction, to avoid the over fitting problem, early training-terminating method called validation was employed. In this work, generated data were divided into three parts; training, validation and testing. The best MLP was selected based on one with the smallest test error. MLPs were trained on randomized data for enhancing learning capability. Basic work stages are shown in the flow chart Figure 10. The network growing technique was used to obtain an optimal MLP size. Network growing basically adds one hidden node at a time into an ANN.

MLP was trained from one to thirty hidden layer neurons, results support the selected range. Each network with each number of hidden layer neuron was trained thirty times with random initial free parameters. MLP performance was analyzed based on mean square error (MSE) and mean absolute error (MAE). The MLP with the lowest errors out of thirty run was selected for further comparison. Value of error varies in a particular fashion. Initially error value was high but decreasing with increase of hidden layer size, after touching low it starts either increasing or floating. Out of these one to thirty hidden nodes, size of neurocontroller was chosen based on minimum MAE and MSE and it is given in Table 8.

Table 8: ANN and Statistical Regression (SR) output comparison

|  | Features | Model 7 $\Delta V_T$ $\delta$ | Model 8 $\Delta V_q$ $\delta$ |
|---|---|---|---|
| ANN | MAE | 0.277 | 0.245 |
|  | MSE | 0.430 | 0.395 |
|  | HLN | 23 | 12 |
|  | Time | 0.024 | 0.016 |
| SR | MAE | 1.0937 | 0.8428 |
|  | MSE | 15.4445 | 8.0913 |

SR=Statistical Regression; HLN= Hidden Layer Neurons

The ANN output is almost in proportion to statistical regression output. However the difference between errors of different sets is not in proportion to difference using statistical methods. This manifests ANN's ability to efficiently map highly complex functions. Table 8 gives the comparison of ANN output and statistical regression results. It also shows the size of hidden layer at minimum error value.

Model 8 has lower ANN error at lower hidden layer size of twelve neurons than model 7. Table 8 also depicts performance comparison of ANN, trained on both models, based on time. Model 8 has higher prediction accuracy on regression analysis and also has lesser error at lesser processing time. The lower processing time is because of smaller hidden layer size.

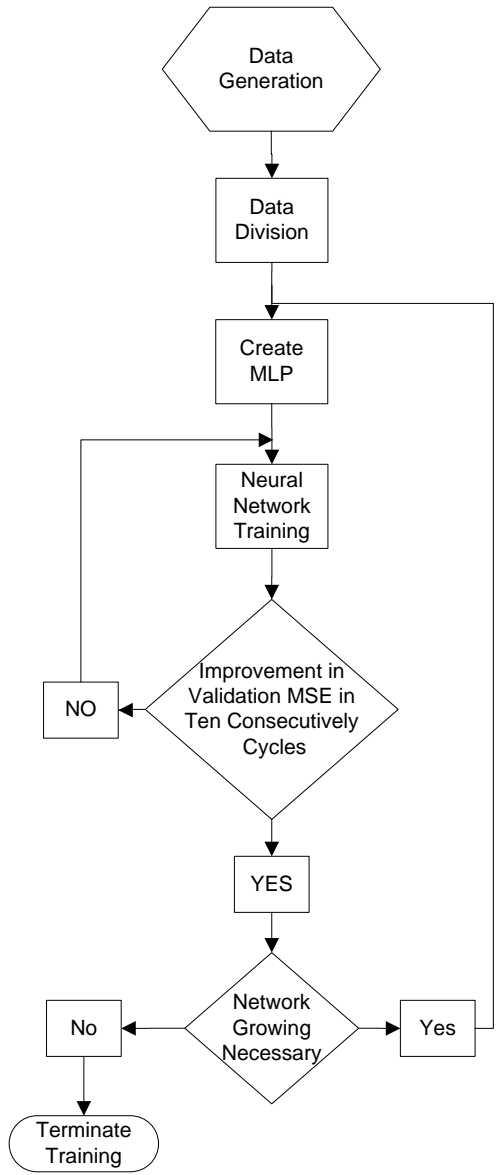

Figure 10 Flow chart showing MLP training stages

## 5. Conclusion

With the help of statistical analysis it is, revealed that strong correlation between input and output space enhance learning capability of ANN not only in terms of error value but also requires lesser hidden layer size. Combination of quadratic voltage and load angle (δ) is a better set of input features for synchronous generator excitation system neurocontroller training. Comparison imparts that the performance of ANN is superior to statistical regression.

## Acknowledgments

The authors greatly acknowledge the support of the Universiti Sains Malaysia fellowship scheme for financial support.

Authors also want to acknowledge Associate Professor Dr Zalina Abdul Aziz for her valuable discussion and guidance from time to time.

Abdul Ghani Abro received his first and second degree from MUET University – Pakistan and NED University – Pakistan. He is Assistant Professor at NED University – Pakistan and currently he is pursuing PhD studies at School of Electrical and Electronic Engineering Universiti Sains Malaysia. His research interests include application of intelligent systems for power system operation and control.

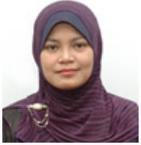

Junita Mohamad-Saleh received her B.Sc (in Computer Engineering) degree from the Case Western Reserve University, USA in 1994, the M.Sc. degree from the University of Sheffield, UK in 1996 and the Ph.D. degree from the University of Leeds, UK in 2002. She is currently an Associate Professor in the School of Electrical & Electronic Engineering, Universiti Sains Malaysia. Her research interests include computational intelligence, tomographic imaging and soft computing